# Acoustic spin pumping: Direct generation of spin currents from sound waves in Pt/Y$_3$Fe$_5$O$_{12}$ hybrid structures

K. Uchida,[1,2,*] H. Adachi,[2,3] T. An,[1,2] H. Nakayama,[1,2] M. Toda,[4] B. Hillebrands,[5] S. Maekawa,[2,3] and E. Saitoh[1,2,3]

[1]*Institute for Materials Research, Tohoku University, Sendai 980-8577, Japan*
[2]*CREST, Japan Science and Technology Agency, Sanbancho, Tokyo 102-0075, Japan*
[3]*Advanced Science Research Center, Japan Atomic Energy Agency, Tokai 319-1195, Japan*
[4]*Graduate School of Engineering, Tohoku University, Sendai 980-8579, Japan*
[5]*Fachbereich Physik and Forschungszentrum OPTIMAS,
Technische Universität Kaiserslautern, 67663 Kaiserslautern, Germany*

Using a Pt/Y$_3$Fe$_5$O$_{12}$ (YIG) hybrid structure attached to a piezoelectric actuator, we demonstrate the generation of spin currents from sound waves. This "acoustic spin pumping" (ASP) is caused by the sound wave generated by the piezoelectric actuator, which then modulates the distribution function of magnons in the YIG layer and results in a pure-spin-current injection into the Pt layer across the Pt/YIG interface. In the Pt layer, this injected spin current is converted into an electric voltage due to the inverse spin-Hall effect (ISHE). The ISHE voltage induced by the ASP is detected by measuring voltage in the Pt layer at the piezoelectric resonance frequency of the actuator coupled with the Pt/YIG system. The frequency-dependent measurements enable us to separate the ASP-induced signals from extrinsic heating effects. Our model calculation based on the linear response theory provides us with a qualitative and quantitative understanding of the ASP in the Pt/YIG system.

## I. INTRODUCTION

A spin pumping effect refers to the transfer of spin-angular momentum from magnetization dynamics in a ferromagnet to conduction-electron spins in an attached paramagnet;[1–14] when a magnetization motion in the ferromagnet is excited, a spin current[11,15,16] is pumped out of the ferromagnet into the paramagnet. The spin pumping is of immense importance in spintronics[17–22] since it allows the generation of pure spin currents using a simple ferromagnetic/paramagnetic bilayer structure.

The conventional spin pumping has been operated by microwaves of several GHz frequencies under ferromagnetic resonance (FMR)[1–8,10,12–14] or spin-wave resonance (SWR)[9,11] conditions, where a steady magnetization motion is maintained in a ferromagnet. This microwave-driven spin pumping has been observed in a variety of sample systems ranging from magnetic metals and semiconductors to magnetic insulators covered with paramagnets.

In a recent letter, we have demonstrated a new type of spin pumping: *acoustic* spin pumping (ASP).[23] We showed that spin currents are generated by an external injection of sound waves into a ferromagnet and that the sound-wave-driven spin currents can be detected electrically by means of the inverse spin-Hall effect[5–14,24–26] (ISHE) in a paramagnetic metal attached to the ferromagnet. The ASP has been operated by sound waves of several MHz frequencies ($<10$ MHz, far below the FMR or SWR frequencies) as a consequence of the energy transfer from sound waves to spin waves, or magnons, in a ferromagnet. In this paper, we report systematic experiments and model calculations on the ASP in paramagnetic metal (Pt)/ferrimagnetic insulator (Y$_3$Fe$_5$O$_{12}$: YIG) hybrid structures.

## II. EXPERIMENTAL PROCEDURE

Figure 1(a) shows a schematic illustration of the sample system used in the present study. The sample con-sists of a single-crystal YIG slab with a Pt film attached to the (100) surface of the YIG. The YIG surface was well polished with alumina paste, and the resultant surface roughness was $< 2$ nm. The Pt layer was fabricated by an RF magnetron sputtering. The lengths of the YIG slab (Pt film) along the $x$, $y$, and $z$ directions are, respectively, 6 mm ($L_{\text{Pt}} = 6$ mm), 2 mm (0.5 mm), and $d_{\text{YIG}} = 1$ mm (15 nm), except when collecting a set of sample-size-dependent data shown in Sec. III C. The Pt/YIG sample is covered with a silicone-rubber heat sink and is fixed on a piezoelectric actuator: a polyvinylidene-fluoride (PVDF) film of the thickness of 52 $\mu$m or a lead-zirconate-titanate (Pb(Zr,Ti)O$_3$: PZT) ceramic of the thickness of $d_{\text{PZT}}$ (=0.6 mm, 0.4 mm, and 0.3 mm) (see Fig. 1(a)). Each actuator is of $10 \times 3$ mm$^2$ rectangular shape. The PVDF film and the PZT ceramic were used for the experiments in Secs. III A-III C and for the experiments in Sec. III D, respectively. When an AC voltage is applied between the top and bottom electrodes of the actuator, it vibrates in the thickness direction ($z$ direction) and injects longitudinal sound waves into the attached YIG slab (see Fig. 1).

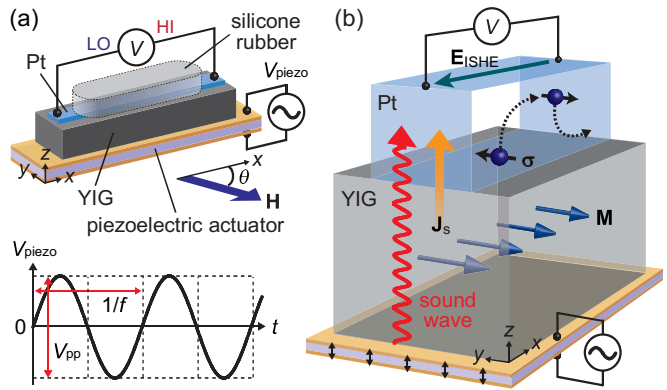

FIG. 1: (a) A schematic illustration of the Pt/YIG sample. The sample was attached to a piezoelectric PVDF film (Measurement Specialties) or a PZT ceramic (Fuji Ceramics C-92H) with silver paste. The AC voltage $V_{\text{piezo}}$ with the frequency $f$ and the peak-to-peak voltage $V_{\text{pp}}$ was applied to the piezoelectric actuator by using a multifunction synthesizer (NF WF1946B). $\theta$ is the angle between the magnetic field **H** and the $x$ direction. The DC voltage $V$ between the ends of the Pt layer was measured with a nanovoltmeter (Keithley 2182A). (b) The acoustic spin pumping (ASP) and the inverse spin-Hall effect (ISHE) in the Pt/YIG sample. $\mathbf{E}_{\text{ISHE}}$, $\boldsymbol{\sigma}$, $\mathbf{J}_s$, and $\mathbf{M}$ denote the electric field generated by the ISHE, the spin-polarization vector of electrons in the Pt, the spatial direction of the ASP-induced spin current, and the magnetization vector of the YIG, respectively.

The mechanism of the measurement is as follows. In the Pt/YIG/piezoelectric-actuator system, if the injected sound wave modulates the magnon distribution function in the YIG slab via the magnon-phonon interaction, it induces a spin current[23] with the spatial direction $\mathbf{J}_s$ and the spin-polarization vector $\boldsymbol{\sigma}$ parallel to the magnetization $\mathbf{M}$ of the YIG slab in the Pt layer, since the magnetic moments in the YIG and conduction-electrons' spins in the Pt are coupled via the interface spin exchange.[11] This sound-wave-driven spin current is converted into a DC electric field $\mathbf{E}_{\text{ISHE}}$ due to the ISHE in the Pt layer owing to the strong spin-orbit interaction in the Pt.[10,13,25] When $\mathbf{M}$ of the YIG slab ($\parallel \boldsymbol{\sigma}$ in the Pt film) is along the $y$ direction ($\theta = 90°$), $\mathbf{E}_{\text{ISHE}}$ is generated along the $x$ direction because of the following ISHE symmetry (see Fig. 1(b)):[5,7,13]

$$\mathbf{E}_{\text{ISHE}} \propto \mathbf{J}_s \times \boldsymbol{\sigma}. \quad (1)$$

Therefore, by measuring $\mathbf{E}_{\text{ISHE}}$, we can detect the ASP electrically. Notable is that, in this setup, we can extract the pure contribution of the magnon-phonon interaction since YIG is an insulator, in which extrinsic artifacts due to charge currents and short-circuit effects[13] are completely excluded. To detect the ASP-induced ISHE signal, we measured a DC electric voltage difference $V$ between the ends of the Pt layer with applying an AC voltage with the frequency $f$ and the peak-to-peak voltage $V_{\text{pp}}$ to the actuator. During the measurements, an external magnetic field $\mathbf{H}$ with the magnitude $H$ was applied at an angle $\theta$ to the $x$ direction (see Fig. 1(a)). All the measurements were performed at room temperature and atmospheric pressure.

## III. RESULTS AND DISCUSSION

### A. ASP in Pt/YIG/PVDF sample systems

Figures 2(a) and 2(b) respectively show the $f$ dependences of the DC voltage $V$ between the ends of the Pt layer in the Pt/YIG/PVDF sample at $H = 1$ kOe and $\theta = 90°$ and the temperature rise of the PVDF film $\Delta T_{\text{PVDF}}$, measured with applying the AC voltage of $V_{\text{pp}} = 10$ V to the PVDF film. When $f > 5$ MHz, $\Delta T_{\text{PVDF}}$ monotonically increases with increasing $f$ (see Fig. 2(b)). In this frequency range, $V$ signals of positive sign appear and the shape of the $f$-$V$ curve is similar to that of the $f$-$\Delta T_{\text{PVDF}}$ curve (compare Figs. 2(a) and 2(b)), indicating that this background of positive sign in the $V$ spectrum is attributed to the heating of the PVDF film, i.e., the conventional spin-Seebeck effect (SSE),[27–37] consistent with the previous experiments on the longitudinal SSE in Pt/YIG systems.[32] In contrast, in the $V$ spectrum, a sharp dip structure of negative sign was found to appear around $f = 3.5$ MHz, although $\Delta T_{\text{PVDF}}$ is negligibly small in this frequency range (see Figs. 2(a) and 2(b)). Because of the different sign, this dip structure is clearly irrelevant to the SSE induced by the heating of the PVDF film (see also Sec. III E). We confirmed that the magnitude of the observed $V$ dip is proportional to the electric power applied to the PVDF film ($\propto V_{\text{pp}}^2$), i.e., the sound-wave intensity in the YIG slab (see Figs. 2(c) and 2(d)). Since the frequency of the $V$ dip ($f = 3.5$ MHz) is far below the FMR and SWR frequencies in YIG ($\sim$ GHz),[11] the signal observed here is also irrelevant to conventional elastically-driven magnetic resonance.[38–40]

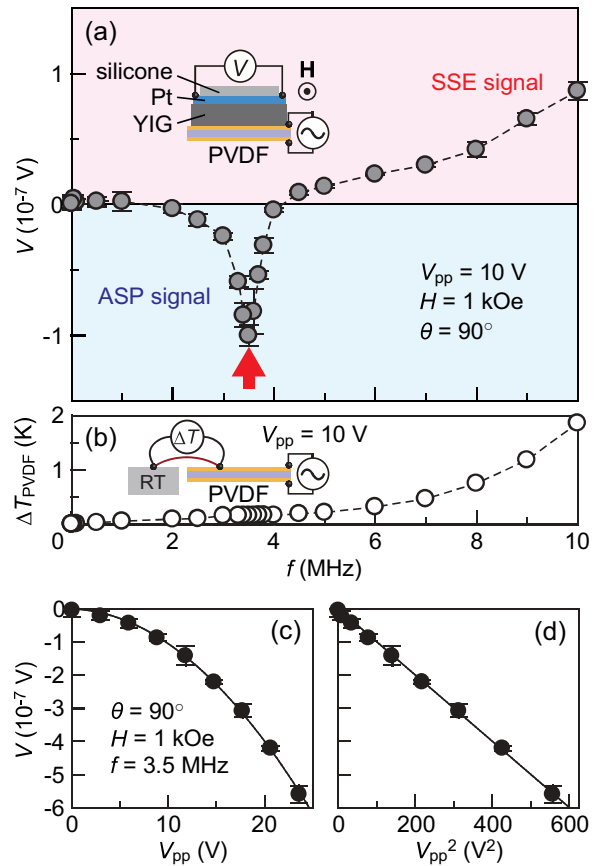

FIG. 2: (a) $f$ dependence of $V$ between the end of the Pt layer in the Pt/YIG/PVDF sample at $V_{\text{pp}} = 10$ V, $H = 1$ kOe, and $\theta = 90°$. As shown in Sec. III A, the negative (positive) $V$ signal is due to the ASP (spin-Seebeck effect: SSE). (b) $f$ dependence of $\Delta T_{\text{PVDF}}$ at $V_{\text{pp}} = 10$ V. Here, $\Delta T_{\text{PVDF}}$ denotes the temperature rise of the PVDF film, due to the applied AC voltage, measured with a differential thermocouple attached between the PVDF and a heat bath at room temperature (RT). (c), (d) $V_{\text{pp}}$ and $V_{\text{pp}}^2$ dependences of $V$ in the Pt/YIG/PVDF sample at $f = 3.5$ MHz, $H = 1$ kOe, and $\theta = 90°$. The error bars represent 95 % confidence level.

Figure 3(a) shows the $H$ dependence of $V$ in the Pt/YIG sample for various values of $f$. When the magnetic field $\mathbf{H}$ is applied along the $\theta = 90°$ direction, the sign of $V$ at the dip position is reversed by reversing $H$. We also checked that the $V$ signal disappears in the Pt/YIG/PVDF sample at $\theta = 0$ (Fig. 3(b)) and in a Cu/YIG/PVDF sample (Fig. 3(c)), where the Pt layer is replaced with a same-sized Cu film with weak spin-orbit interaction, a situation consistent with the feature of the ISHE described by Eq. (1). All the results shown above support that the $V$ dip around $f = 3.5$ MHz observed in the Pt/YIG/PVDF sample is attributed to the spin currents driven by the ASP. The experimental data shown in Figs. 2(a) and 3(a)-3(c) are summarized in the contour plots in Figs. 3(d)-3(f).

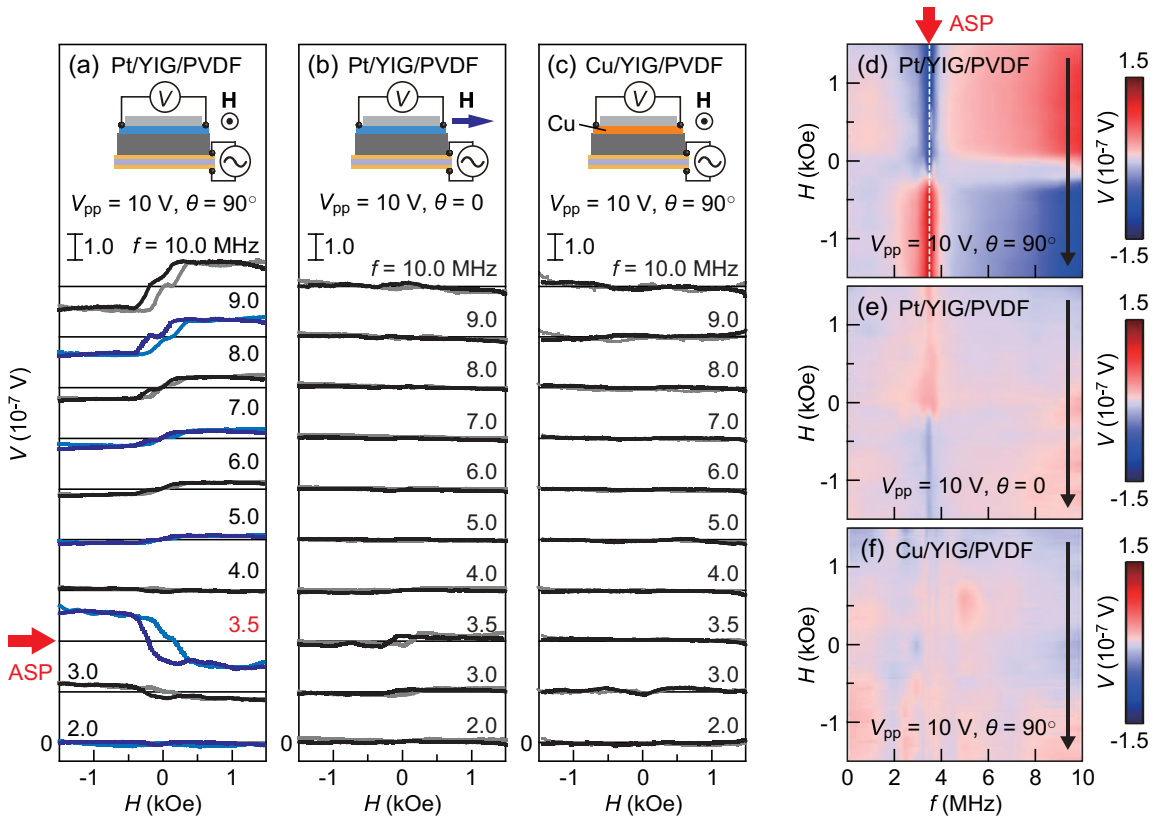

FIG. 3: (a), (b) $H$ dependence of $V$ in the Pt/YIG/PVDF sample for various values of $f$ at $V_{\rm pp} = 10$ V and $\theta = 90°$ (a) or $\theta = 0$ (b). (c) $H$ dependence of $V$ in the Cu/YIG/PVDF sample for various values of $f$ at $V_{\rm pp} = 10$ V and $\theta = 90°$. (d), (e) Contour plot of $V$ in the Pt/YIG/PVDF sample as functions of $f$ and $H$ at $V_{\rm pp} = 10$ V and $\theta = 90°$ (d) or $\theta = 0$ (e). (f) Contour plot of $V$ in the Cu/YIG/PVDF sample as functions of $f$ and $H$ at $V_{\rm pp} = 10$ V and $\theta = 90°$. The black arrows represent the $H$-sweep directions.

### B. Effect of silicone rubber on the device

In this subsection, we discuss $V$ signals measured when a silicone rubber was removed from the Pt/YIG sample. Figure 4(a) shows $V$ as a function of $f$ in the same Pt/YIG/PVDF sample at $V_{\rm pp} = 10$ V, measured with and without the silicone rubber. Here, the magnetic field of 1 kOe was applied along the $\theta = 90°$ direction. In the absence of the silicone rubber, the SSE signal in the higher-frequency range ($f > 5$ MHz) is suppressed, which is attributed to the very low thermal conductivity of air that reduces the temperature gradient across the Pt/YIG sample, or the SSE signal. Nevertheless, the $V$ dip around $f = 3.5$ MHz is almost unchanged even in the sample without the silicone rubber. This result indicates that the ISHE voltage induced by the ASP does not depend on the direction of injected sound waves, since most of sound waves are reflected from the top surface of the sample because of the mismatch of the characteristic acoustic impedance[41] between the Pt/YIG sample and air. Therefore, the ASP is attributed not to the momentum transfer or Doppler shift,[42] but to the energy transfer from sound waves to magnons in the YIG slab. This interpretation is formulated in terms of a model calculation shown in Sec. III F.

### C. Sample-size dependence

Figure 4(b) shows the dependence of the $f$-$V$ curves on the length of the Pt layer, $L_{\rm Pt}$, in the Pt/YIG/PVDF samples at $V_{\rm pp} = 10$ V, $H = 1$ kOe, and $\theta = 90°$. The observed $V$ signal at $L_{\rm Pt} = 2$ mm is much smaller than that at $L_{\rm Pt} = 6$ mm; the ISHE voltage is proportional to $L_{\rm Pt}$. Notable is that the position of the $V$ dip does not change even in the Pt/YIG/PVDF sample of $L_{\rm Pt} = 2$ mm, confirming that the shape effect of the Pt film, such as a mechanical resonance of the Pt layer, is irrelevant to this $V$-dip generation.

We also confirmed that the shape of the $f$-$V$ curve and the $V$-dip position do not change in Pt/YIG/PVDF samples of different YIG-slab thicknesses (see the inset to Fig. 4(b)). This result indicates that the $V$-dip signal is not induced from standing sound waves; the multiple reflection and interference of sound waves between the top and bottom surfaces of the YIG slab are also irrelevant.

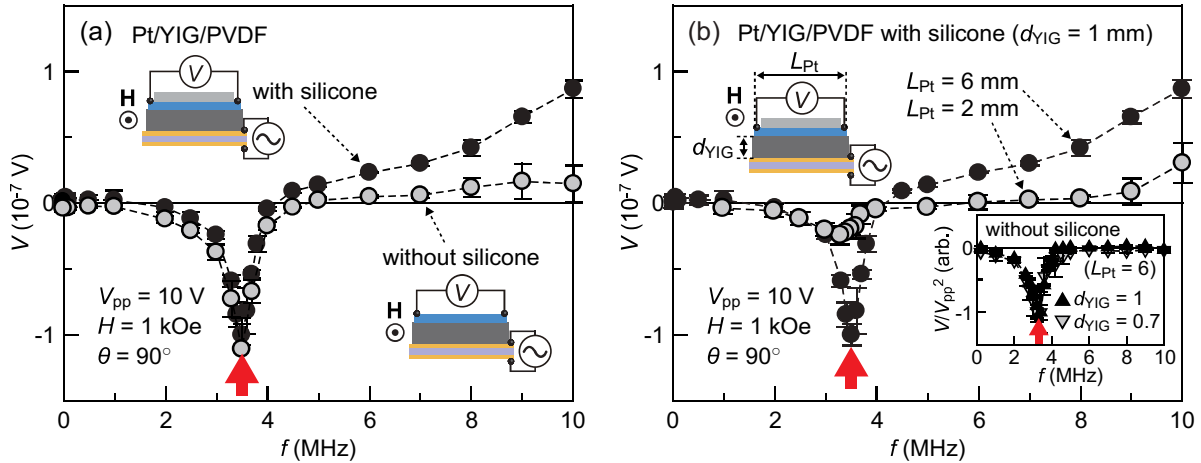

FIG. 4: (a) $f$ dependence of $V$ in the Pt/YIG/PVDF sample at $V_{pp} = 10$ V, $H = 1$ kOe, and $\theta = 90°$, measured with and without the silicone rubber. (b) $f$ dependence of $V$ in the Pt/YIG ($d_{YIG} = 1$ mm)/PVDF samples of the two different Pt-film lengths ($L_{Pt} = 6$ mm and 2 mm) at $V_{pp} = 10$ V, $H = 1$ kOe, and $\theta = 90°$. The inset to (b) shows that the $f$ dependence of the normalized $V/V_{pp}^2$ in the Pt ($L_{Pt} = 6$ mm)/YIG/PVDF samples of the two different YIG-slab thicknesses ($d_{YIG} = 1$ mm and 0.7 mm) at $H = 1$ kOe and $\theta = 90°$.

### D. Piezoelectric-resonance-frequency dependence

Next, we demonstrate that the sharp $V$-dip structure generated by the ASP is due to the piezoelectric resonance of the actuator attached to the Pt/YIG sample. To do this, we used piezoelectric PZT ceramics with different thicknesses ($d_{PZT}$) and different piezoelectric-resonance frequencies ($f_p$), instead of the PVDF film. We confirmed that the PZTs of $d_{PZT} = 0.6$ mm, 0.4 mm, and 0.3 mm exhibit the piezoelectric resonance at $f = 3.6$ MHz, 5.4 MHz, and 7.2 MHz, respectively, by means of a laser Doppler vibrometry (see Figs. 5(b) and 5(c)).

Figure 5(a) shows the $f$-$V$ curves for the Pt/YIG/PZT samples for various values of $f_p$ (= 3.6 MHz, 5.4 MHz, and 7.2 MHz) at $V_{pp} = 10$ V, $H = 1$ kOe, and $\theta = 90°$, in which the $f_p$ positions are marked with arrows. In all the Pt/YIG/PZT samples, the $V$ dip was found to appear around $f = f_p$. These $V$ signals are dominated by the ASP since the sign of $V$ is negative, opposite to the SSE signals induced by the heating of the PZTs (see Sec. III E). This result confirms that the sharp $V$ structure of negative sign originates from the resonant sound-wave injection generated by the piezoelectric resonance of the actuator, not the shape effects of the Pt/YIG samples.

As an endnote to this subsection, we mention the difference in the magnitudes of the ASP signals observed in the Pt/YIG/PZT and in the Pt/YIG/PVDF samples. As shown in Figs. 2, 3, and 5, the $V$-dip signals in the Pt/YIG/PZT samples are one order of magnitude greater than those in the Pt/YIG/PVDF samples. This is because (1) the piezoelectric strain constant of the PZT ceramic ($d_{33} = 770 \times 10^{-12}$ m/V for thickness mode) is much greater than that of the PVDF film ($d_{33} = -33 \times 10^{-12}$ m/V) and (2) the energy transmittance of sound waves at the YIG/PZT interface is about 3 times greater than that at the YIG/PVDF interface.[43] The relatively-small $V$-dip signal in the Pt/YIG/PZT sample of $d_{PZT} = 0.3$ mm can be explained by the large temperature rise of the PZT of $d_{PZT} = 0.3$ mm around $f_p = 7.2$ MHz (see Fig. 5(e)), where the ASP signal of negative sign is offset by the SSE background signal of positive sign.

### E. Sign difference between ASP and SSE

Using Fig. 6, we discuss qualitatively the origin of the sign difference between the signal coming from the ASP and that from the SSE induced by the heating of the piezoelectric actuator (see Fig. 2(a)). At the Pt/YIG interface, there are two contributions to the spin injection process: the excitations of magnons in the YIG slab which *inject* spin currents into the Pt film and the excitations of conduction electrons in the Pt film which *eject* spin currents from itself.[35] Then, in the case of the sound-wave-driven spin injection, sound waves interact with magnons in the YIG slab efficiently, while they do not efficiently interact with conduction electrons in the Pt film because the thickness of the Pt film (15 nm) is too small for conduction electrons in the Pt film to feel the sound waves of several MHz frequencies which have wavelengths of the order of millimeters. This means that the sound waves do efficiently excite magnons in YIG slab while not conduction electrons in the Pt film (see Figs. 6(a) and 6(b)); spin currents are injected into the Pt film. On the other hand, in the case of heating by the piezoelectric actuator, magnon densities in the YIG slab and electrons in the Pt layers are excited by heat currents, i.e., flows of thermal phonons. Since the thermal de Broglie length of phonons is much shorter than a millimeter, thermal phonons, of which the densest frequency at 300 K is $\sim 20$ THz,[44] are able to excite conduction electrons in the Pt film. Moreover, since the silicone rubber in the present setup acts as a heat sink well, the heat currents carried by thermal phonons penetrate into the Pt film. Under such circumstances, thus, conduction electrons in the Pt film are excited much stronger than magnons in the YIG slab (see Figs. 6(c) and 6(d)) as the electron-phonon interaction in the Pt film is much stronger than the magnon-phonon interaction in the YIG slab.[32,45,46] This means that, in the latter case, spin cur-rents are *ejected* from the Pt film as was demonstrated in recent experiments.[32] These considerations explain the difference in the sign of the signals for the ASP and SSE processes as seen in Fig. 2(a).

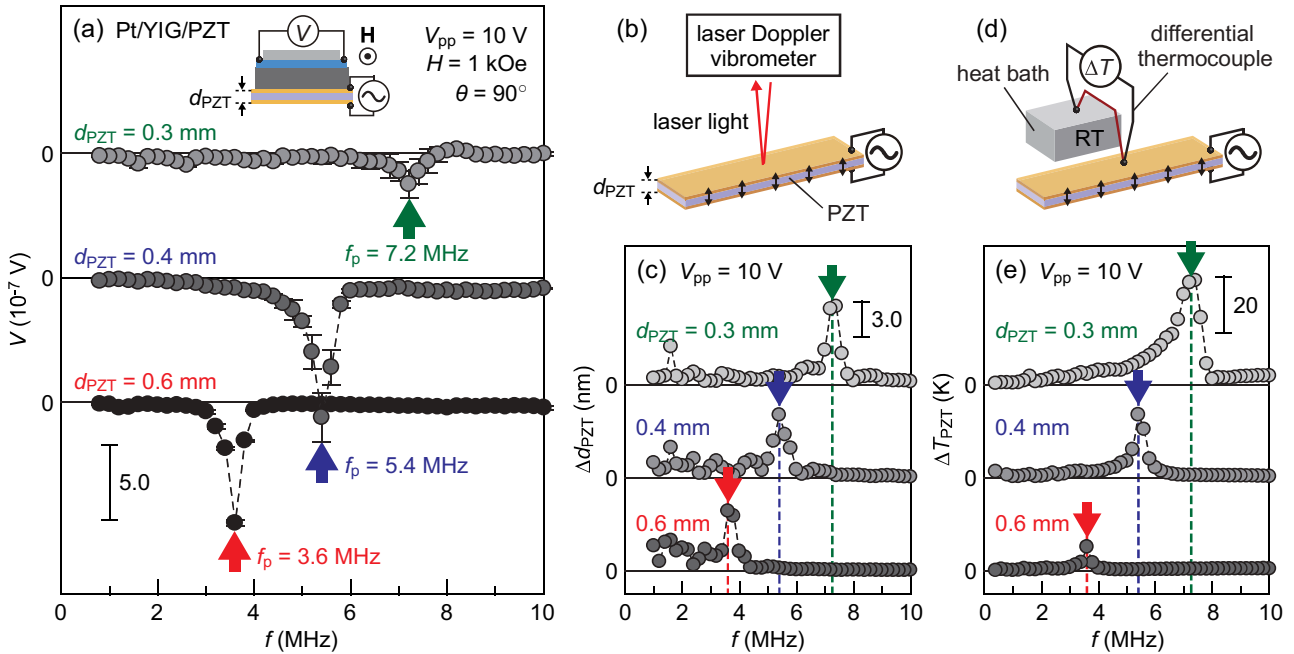

FIG. 5: (a) $f$ dependence of $V$ in the Pt/YIG/PZT samples for various PZT thicknesses ($d_{\rm PZT} = 0.6$ mm, 0.4 mm, and 0.3 mm) at $V_{\rm pp} = 10$ V, $H = 1$ kOe, and $\theta = 90°$. (b) Experimental configuration for the measurement of the thickness-vibration amplitude of the PZT ceramics ($\Delta d_{\rm PZT}$), due to the applied AC voltage. $\Delta d_{\rm PZT}$ was measured by means of a laser Doppler vibrometry with a microsystem analyzer (Polytec MSA-500). (c) $f$ dependence of $\Delta d_{\rm PZT}$ for various values of $d_{\rm PZT}$ at $V_{\rm pp} = 10$ V. The peak position of each $f$-$\Delta d_{\rm PZT}$ curve corresponds to the piezoelectric resonance frequency $f_{\rm p}$ of each PZT. (d) Experimental configuration for the measurement of the temperature rise of the PZT ceramics ($\Delta T_{\rm PZT}$), due to the applied AC voltage. $\Delta T_{\rm PZT}$ was measured with a differential thermocouple attached between the PZT and a heat bath at RT. (e) $f$ dependence of $\Delta T_{\rm PZT}$ for various values of $d_{\rm PZT}$ at $V_{\rm pp} = 10$ V.

### F. Model calculation using linear-response theory

In this subsection, we present a linear-response approach to the ASP and show that the experimental observation can be explained as the energy transfer from sound waves to magnons. As discussed in Sec. III E, our task reduces to calculating the Feynman diagram shown in Fig. 6(a), where the magnons in the YIG slab are excited by external sound waves and, at the Pt/YIG interface, the magnons in the YIG and the itinerant spin density in the Pt interact weakly through the s-d exchange coupling near the interface. As in Ref. 23, we consider here the interaction of exchange origin between magnons and sound waves (the so-called volume magnetostrictive coupling[47]), since this coupling has been established to give the largest contribution.[48] We neglect the so-called single-ion magnetostriction[47] arising from the spin-orbit interaction;[49] if the latter coupling was relevant to our experiment, the resultant ASP should have be seen at the GHz frequencies, because the single-ion magnetostriction contains the coupling linear in the transverse magnetic fluctuations $S_x$ and $S_y$,[49] thereby converting phonons into magnons with the same frequency as in the case of FMR, where photons are converted into magnons with the same frequency.

The starting model is the following exchange Hamiltonian:

$$\mathcal{H}_{\rm ex} = -\sum_{\bm{R}_i, \bm{R}_j} J_{\rm ex}(\bm{R}_i - \bm{R}_j)\, \bm{S}(\bm{R}_i) \cdot \bm{S}(\bm{R}_j), \quad (2)$$

where $J_{\rm ex}(\bm{R}_i - \bm{R}_j)$ is the strength of the exchange coupling between the ions at $\bm{R}_i$ and $\bm{R}_j$. The instantaneous position of the ion is written as $\bm{R}_i = \bm{r}_i + \bm{u}(\bm{r}_i)$, where the lattice displacement $\bm{u}(\bm{r}_i)$ is separated from the equilibrium position $\bm{r}_i$. In the absence of sound waves ($\bm{u} = \bm{0}$), the exchange Hamiltonian can be diagonalized by introducing the magnon operator $a(\bm{r}_i)$ and $a^\dagger(\bm{r}_i)$ in the following manner: $S^+(\bm{r}_i) \equiv (1/\sqrt{2})(S^x + {\rm i}S^y) = \sqrt{S_0}a^\dagger(\bm{r}_i)$, $S^-(\bm{r}_i) \equiv (1/\sqrt{2})(S^x - {\rm i}S^y) = \sqrt{S_0}a(\bm{r}_i)$, and $S^z(\bm{r}_i) = -S_0 + a^\dagger(\bm{r}_i)a(\bm{r}_i)$. This yields the magnon Hamiltonian

$$\mathcal{H}_{\rm mag} = \sum_{\bm{q}} \omega_{\bm{q}} a^\dagger_{\bm{q}} a_{\bm{q}}, \quad (3)$$

where $\omega_{\bm{q}} = 2S_0 \sum_{\bm{\delta}} J_{\rm ex}(\bm{\delta}) \sum_{\bm{q}} \left[1 - \cos(\bm{q} \cdot \bm{\delta})\right]$ is the magnon frequency with the lattice vector $\bm{\delta}$ for the nearest neighbors.

Now we consider the effect of sound waves. In the presence of sound waves, the exchange Hamiltonian can be written as $\mathcal{H}_{\rm ex} = \mathcal{H}_{\rm mag} + \mathcal{H}_{\rm mag-sound}$ with $\mathcal{H}_{\rm mag-sound}$ being the interaction between magnons and sound waves. When the sound waves propagate along the symmetry axis of the crystal (this is the case for the present experiment, see Sec. sec:II), $\mathcal{H}_{\rm mag-sound}$ is approximately given by

$$\mathcal{H}_{\rm mag-sound} \approx \widetilde{g} \sum_{\bm{r}_i, \bm{\delta}} \left(\bm{\nabla} \cdot \bm{u}(\bm{r}_i)\right) a^\dagger(\bm{r}_i) a(\bm{r}_i + \bm{\delta}) \quad (4)$$

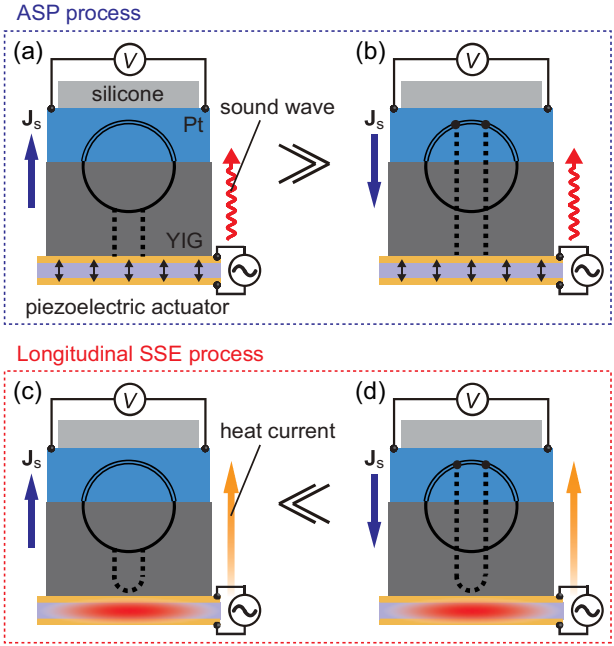

FIG. 6: (a), (b) Feynman diagrams for calculating the ASP-induced spin currents in the Pt layer in the Pt/YIG sample. In the case of the ASP, the process in (a) is dominant. (c), (d) Feynman diagrams for calculating the longitudinal-SSE-induced spin currents in the Pt layer in the Pt/YIG sample. In the case of the longitudinal SSE, the process in (d) is dominant. The double lines, bold lines, and dotted lines represent spin-density propagators, magnon propagators, and external sound waves (in (a) and (b)) or heat currents (in (c) and (d)), respectively.

up to the linear order in the displacement field, where $\widetilde{g} = (\sum_{\boldsymbol{\delta}} \boldsymbol{\delta} \cdot \nabla J_{\text{ex}}(\boldsymbol{\delta}))/2J_{\text{ex}}(\boldsymbol{\delta})$ is the dimensionless phonon-magnon coupling constant. Equation (4) means that, up to the lowest order in $\boldsymbol{u}$, only the longitudinal sound waves couple to magnons when the sound waves propagate along the symmetry axis of the crystal.[48] We express the lattice displacement field $\boldsymbol{u}$ for a fixed wavenumber $\boldsymbol{K}_0$ as $\boldsymbol{u}(\boldsymbol{r}_i) = i \sum_{\boldsymbol{K}=\pm\boldsymbol{K}_0} \widehat{\boldsymbol{e}}_{\boldsymbol{K}} U_{\boldsymbol{K}} e^{i\boldsymbol{K}_0 \cdot \boldsymbol{r}_i}$,[50] where the polarization vector $\widehat{\boldsymbol{e}}_{\boldsymbol{K}}$ is odd under the inversion $\boldsymbol{K} \to -\boldsymbol{K}$, and $U_{\boldsymbol{K}}$ can be expressed as $U_{\boldsymbol{K}} = u_{\boldsymbol{K}} + u^*_{-\boldsymbol{K}}$ to satisfy $U_{\boldsymbol{K}} = U^*_{-\boldsymbol{K}}$. Note that the spatial average of $[\boldsymbol{u}(\boldsymbol{r}_i)]^2$ is given by $\langle [\boldsymbol{u}(\boldsymbol{r}_i)]^2 \rangle_{\text{av}} = 2|U_{\boldsymbol{K}_0}|^2$. Then, going into the momentum representation, $\mathcal{H}_{\text{mag-sound}}$ is written as

$$\mathcal{H}_{\text{mag-sound}} = \widetilde{g} \sum_{\boldsymbol{q},\boldsymbol{K}=\pm\boldsymbol{K}_0} (\hbar\omega_{\boldsymbol{q}})(\boldsymbol{K}\cdot\widehat{\boldsymbol{e}}_{\boldsymbol{K}}) U_{\boldsymbol{K}} a^\dagger_{\boldsymbol{q}+\boldsymbol{K}} a_{\boldsymbol{q}}. \quad (5)$$

Here, we calculate the spin current $J_s$ injected into the Pt film. This quantity is defined by the statistical average of the rate of change of the spin density $\boldsymbol{s}$ in the Pt film as $J_s = \sum_{\boldsymbol{r}\in\text{Pt}} \langle \partial_t s^z(\boldsymbol{r},t) \rangle$. As is inferred from the similarity between Eq. (5) above and Eq. (5) in Supplementary Information of Ref. 23, the following calculation is mostly the same as that given in our previous works[23,33] except for one important difference that the phonon field in the previous calculation is a statistical variable obeying the Bose statistics, while the phonon field in the present case plays a role of an external field. In the Feynman diagram relevant to the ASP (Fig. 6(a)), the double line, bold line, and dotted lines represent a spin-density propagator in the Pt film, a magnon propagator in the YIG slab, and external sound waves, respectively. Repeating essentially the same procedure as in our previous works,[23,33] the spin current injected into the Pt film $J_s$ by the process shown in Fig. 6(a) is then calculated as

$$J_s = \frac{\sqrt{2}\hbar(J_{\text{sd}}^2 S_0)}{N_P N_F / N_{\text{int}}} \sum_{\boldsymbol{k},\boldsymbol{q}} A_{\boldsymbol{k},\boldsymbol{q}}(\nu_{\boldsymbol{K}_0}) \Big(\widetilde{g}(\hbar\omega_{\boldsymbol{q}})|\boldsymbol{K}_0|\Big)^2 |U_{\boldsymbol{K}_0}|^2, \quad (6)$$

where $J_{\text{sd}}$ is the s-d exchange coupling at the Pt/YIG interface, $S_0$ the size of the localized spin in YIG, $N_{\text{int}}$ the number of localized spins in YIG, $N_P$ ($N_F$) the number of lattice sites in Pt (YIG). In Eq. (6), the quantity $A_{\boldsymbol{k},\boldsymbol{q}}(\nu)$ is defined by

$$A_{\boldsymbol{k},\boldsymbol{q}}(\nu) = \int \frac{d\omega}{2\pi} \, \text{Im}\chi^R_{\boldsymbol{k}}(\omega) \text{Im} X^R_{\boldsymbol{q}-\boldsymbol{K}}(\omega-\nu) |X^R_{\boldsymbol{q}}(\omega)|^2 \\ \times [\coth(\frac{\hbar(\omega-\nu)}{2k_B T}) - \coth(\frac{\hbar\omega}{2k_B T})], \quad (7)$$

where[35] $\chi^R_{\boldsymbol{k}}(\omega) = \chi_P/(1 + \lambda_{\text{sf}}^2 k^2 - i\omega\tau_{\text{sf}})$ is the retarded component of the spin-density propagator in Pt with $\chi_P$, $\lambda_{\text{sf}}$, and $\tau_{\text{sf}}$ being respectively the paramagnetic susceptibility, the spin-diffusion length, and the spin-relaxation time and $X^R_{\boldsymbol{q}}(\omega) = (\omega - \widetilde{\omega}_{\boldsymbol{q}} + i\alpha\omega)^{-1}$ is the retarded component of the magnon propagator with $\widetilde{\omega}_{\boldsymbol{q}} = \gamma H_0 + \omega_{\boldsymbol{q}}$ and $\alpha$ being respectively the magnon frequency and the Gilbert damping constant. Using $A_{\boldsymbol{k},\boldsymbol{q}}(\nu) \approx -(\chi_P \omega_{\boldsymbol{q}} \tau_{\text{sf}}) \coth(\frac{\omega_{\boldsymbol{q}}}{2T_1})[\frac{1}{\omega_{\boldsymbol{q}}} \text{Im} \frac{\chi_{\boldsymbol{k}}(\omega_{\boldsymbol{q}})}{\chi_P}]^2$, we obtain an approximate expression for the spin current injected into the Pt film as

$$J_s \approx -5 \times 10^{-3} \left( \frac{J_{\text{sd}}^2 S_0 N_{\text{int}} \chi_P}{(\lambda_{\text{sf}}/a)^3} \right) \mathcal{B}_2 (\widetilde{g} K_0 |U_{\boldsymbol{K}_0}|)^2, (8)$$

where $a$ is the lattice spacing in Pt and $\mathcal{B}_2 = \frac{(T/T_m)^{9/2}}{4\pi^2}(\frac{k_B T_m \tau_{\text{sf}}}{\hbar})^3 \int_0^{T_m/T} dv \frac{v^{7/2}}{\tanh(v/2)}$ with the characteristic energy corresponding to the magnon high-energy cutoff $T_m$.

As an order of magnitude estimation, we compare Eq. (8) with the experimental results. When comparing our theoretical and experimental results, a care is necessary because in our model calculation we use effective block spins defined on a cubic lattice with lattice spacing $|\boldsymbol{\delta}| = 1.24$ nm of YIG, while the distance between the two magnetic ions in YIG ($Fe^{3+}$ ions on $a$ site and $d$ site) is 0.34 nm, 3.6 times shorter than the former.[51] Note also that the effective unit cell ($|\boldsymbol{\delta}| = 1.24$ nm) contains 8 formula units. These facts result in a renormalization of the dimensionless magnon-phonon coupling $\widetilde{g}$ defined below Eq. (4), and a naive consideration gives $\widetilde{g} = 3.6 \times 8 \widetilde{g}^{(0)}$ where $\widetilde{g}^{(0)}$ is the bare magnon-phonon coupling constant.

In the experiment, the spin current injected into the Pt film is converted into the electric voltage as $V = \theta_{\text{SH}}(|e|J_s)(\rho/w)$ due to the ISHE (see Eq. (1)), where $\theta_{\text{SH}}$ is the spin-Hall angle, $|e|$ the absolute value of electron charge, $\rho$ the resistivity of the Pt film, and $w$ the width of the Pt layer along the $y$ direction. Using $\lambda_{\text{sf}} = 7$ nm, $a = 0.2$ nm, $|\boldsymbol{\delta}| = 1.24$ nm, $\theta_{\text{SH}} = 0.01$, $\rho = 0.91$ $\mu\Omega$m, $\chi_P = 1 \times 10^{-6}$ cm$^3$/g, $\tau_{\text{sf}} = 1$ ps, $S_0 = 16$, $J_{\text{sd}} = 50$ meV, $T_m = 600$ K, $|U_{\boldsymbol{K}_0}| \sim 5$ nm for $d_{\text{PZT}} = 0.6$ mm at $V_{\text{pp}} = 10$ V, and $v_{\text{sound}} \sim 5$ Km/s for the longitudinal sound velocity of YIG, the experimental signal shown in Fig. 5(a) for $d_{\text{PZT}} = 0.6$ mm can be explained by a value

$\widetilde{g}^{(0)} \sim 1.0$ for the bare magnon-phonon coupling constant. Note that our perturbative approach in terms of the magnon-phonon interaction is ensured by the smallness of $\widetilde{g} \times \text{div}\boldsymbol{u} < 10^{-2}$.

## IV. SUMMARY

In this paper, we report the systematic investigations on the acoustic spin pumping (ASP): the generation of spin currents by a sound-wave injection. Using Pt/$Y_3Fe_5O_{12}$ (YIG)/piezoelectric-actuator hybrid systems, we demonstrate that the sound waves injected in the YIG slab generate a spin current in the Pt film attached to the YIG. The injected spin current is converted into an electric field due to the inverse spin-Hall effect (ISHE) in the Pt film. The ISHE signal induced by the ASP appears at the piezoelectric resonance frequency of the actuator coupled with the Pt/YIG sample. The experimental results show that, in the present Pt/YIG/piezoelectric-actuator systems, the sign of the sound-wave-driven signal is opposite to that of the conventional spin-Seebeck effect due to the heating of the actuator; by checking the signal sign, one can distinguish the ASP-induced signals from extrinsic heating effects. We have developed a linear-response theory for the ASP and shown that it provides us with a qualitative as well as quantitative understanding of the ASP observed in the Pt/YIG samples. Since the ASP enables simple and versatile spin-current generation from sound waves, it will be useful in basic spintronics researches and spin-based device applications.

### Acknowledgements

The authors thank Takahito Ono for valuable discussions. This work was supported by a Grant-in-Aid for Scientific Research A (21244058) from MEXT, Japan, the global COE for the "Materials Integration International Centre of Education and Research" from MEXT, Japan, CREST-JST "Creation of Nanosystems with Novel Functions through Process Integration", Japan, and a Fundamental Research Grant from TUIAREO, Japan.